\documentclass[Proceedings]{ascelike}
\usepackage{subcaption}
\usepackage{amsmath,graphicx,pifont}
\usepackage{epstopdf}
\usepackage{tikz}
\usepackage{pgfplots}
\usepackage{psfrag}
\usetikzlibrary{3d}
\usetikzlibrary{calc}
\usepackage{textcomp}
\usepackage{multirow}
\usepackage{booktabs}
\usepackage{epsfig}
\NameTag{Varadharajan, \today}
\begin{document}
%
\title{Reynolds number effects on transition, turbulence intensity and axial-velocity decay rate of turbulent round jets}
\author{
Ramanathan Varadharajan%
%
\thanks{
Laboratory of Physical Chemistry \& Soft Matter,
Wageningen University \& Soft Matter, 
6708WE Wageningen, The Netherlands.
E-mail: ramanathan.varadharajan@wur.nl},
%
%
 \  
%
%
%
%
}
\maketitle
\begin{abstract}
Numerical simulations of turbulent round jets, using explicit-filtered LES technique, are performed, for three different Reynolds numbers(Re = 3600, 88000, 400000), to understand the Reynolds number effect on subsonic jets with Mach number 0.9. Eigth-order compact schemes are used for spatial derivative estimation. Second order Runge-Kutta method is used for Time evolution of flow. One-parameter fourth-order compact filter is used for low-pass explicit filtering of transported variables. Decreased centreline axial velocity decay rate and reduced turbulence intensities are observed as jet Reynolds number increases. Jet spread-rate is observed to be independent of the Reynolds number. An early potential-core collapse and transition to turbulent regime is observed in jets with high Reynolds number, for same inflow turbulence seeding. Role of increased smaller length scales with increase in Reynolds number is analyzed. It is observed that behavior of turbulent round jets is similar to that of plane jets, irrespective of gain in dimensional freedom. Detailed discussions of the observations made from numerical experiments are presented. 
\end{abstract}
%
%

\section{Introduction}
\label{S:1}

Turbulent round jets have been extensively studied, both experimentally and numerically, for their rich physics and applications.
Applications include but not limited to, turbulent mixing for efficient combustion, industrial whistles, turbulent noise reduction in aircraft jets, etc.,
Thus the evolution of Mean flow and turbulence characteristics of a round jet are extensively researched and documented in literature.
Notable works involve experiments performed by Panchapakesan,\cite{panchapakesan1993turbulence} \& Hussein et al,\cite{hussein1994velocity} to understand mean-flow and turbulence in jets.
Jonathan B. Freund, performed Direct numerical simulations of subsonic round jet at 0.9 Mach Number \cite{freund2001noise} to determine the noise sources in a turbulent jet.

Fellouh \textit{et al.}\cite{fellouah2009reynolds},  proved the Reynolds dependence in mixing transition in near and intermediate-field regions of round jet using hot-wire measurements.
Effect of Reynolds number and initial conditions on turbulent mixing was studied by Boersma\cite{boersma1998numerical},Hussain\cite{hussain1986coherent}, George\cite{george1989self}, Dimotakis\cite{dimotakis2000mixing}, Xu \textit{et al.}\cite{xu2002effect} and Uddin \textit{et al.}\cite{uddin2007self}. A good review of works can be found in Ball \textit{et al.}\cite{ball2012flow}. 
Presence of Reynolds number effects on jet parameters for plane jets was initially reported by Xu \textit{et al.}\cite{xu2013effects}.
Xu's experiments on plane jets shows decrease in centerline decay rates, spread rate with increase in Reynolds number till a critical point is achieved.

In this paper, an extension of the work for round jets is presented, as most of the reported studies on Reynolds number effects were performed on non-circular jets, because of their high mixing efficiency.
It should be noted that the mixing efficiency of circular jets are not insignificant.
It will be shown that for round jets, the dependence on Reynolds number is similar to that of plane jets.
Article also documents the first attempt to study Reynolds number effects on turbulence using explicit-filtered LES technique.
Numerical Methods are explained in Section: \ref{S:2}. An account on the jet inflow conditions used are provided in Section: \ref{S:3}. Computational domain used for various simulations are described in Section: \ref{S:4}. Section: \ref{S:5} presents the results and discussions.

\section{Numerical method}
\label{S:2}

The flows considered is governed by the Navier-Stokes equations for compressible flow
\begin{equation}
\centering
 \frac{\partial \rho}{\partial t} + \frac{\partial \rho u_{i}}{\partial x_{i}} = 0 ,
\end{equation}
\begin{equation}
\centering
\frac{\partial \rho u_{i}}{\partial t} + \frac{\partial \rho u_{i} u_{j}}{\partial x_{j}} = - \frac{\partial p}{\partial x_{i}} + \frac{\partial \tau_{ij}}{\partial x_{j}} ,
\end{equation}
\begin{equation}
\centering
\frac{\partial \rho E}{\partial t} + \frac{\partial}{\partial x_{j}}[(\rho E + p)u_{i}] =  \frac{\partial q_{i}}{\partial x_{i}} + \frac{\partial u_{j} \tau_{ij}}{\partial x_{i}}.
\end{equation}
Here $\rho$ is the density, $u_{i}$ are Cartesian velocity components, $p$ is pressure, $E$ is the energy and $ \rho E = (\rho u_{k} u_{k})/2 + p/(\gamma -1)$. $q_{i}$ is the heat flux.  The Prandtl Number was set to a constant value of 0.71. Shear stress
\begin{equation}
\centering
 \tau_{ij} = \mu \Bigg{(}{\frac{\partial u_{i}}{\partial x_{j}} + \frac{\partial u_{j}}{\partial x_{i}} -\frac{2}{3}\bigg{[}\frac{\partial u_{k}}{\partial x_{k}}\bigg{]}\delta_{ij}}\Bigg{)},
\end{equation}
where $\mu$, the dynamic viscosity, was calculated from Sutherland's law
\begin{equation}
\centering
 \mu = \frac{1.458 \times T^{1.5}}{(T + 110.0)}.
\end{equation}

Since the flows considered are turbulent, the numerical solution was found as a large eddy simulation using the explicit filtering method. Detailed information of the numerical procedure are provided in R. Varadharajan\cite{varadharajan2016study} and S. Ganesh\cite{SubramanianPHD}.  Essential requirements for this method are that a high resolution numerical method be used along with a high resolution low pass spatial filter applied to transported variables after every time step.  This approach to LES has been used successfully, for several types of flows, by at least two other groups \cite{Visbal2002} \cite{Bogey2006}.  Here, a Cartesian grid was used with a 8th-order compact difference formula for spatial derivatives, split into a forward and a backward step extending the method of Hixon and Turkel\cite{hixon1998high}  \cite{hixon2000compact}.  Time-stepping was by a 2nd-order Runge-Kutta(RK2) scheme.  A one-parameter fourth-order compact filter \cite{lele1992compact} was applied with filter paramter $\alpha = 0.475$ in the stream-wise direction and 0.498 in the cross-stream directions.

Partially non-reflecting, characteristic boundary conditions developed by K. W. Thompson\cite{thompson1987time} \cite{thompson1990time} and adapted for compressible fluid flow, by S.K. Lele \cite{poinsot1992boundary}, were specified at boundaries.   A constant time-step estimated from a maximum CFL number of 0.075 was used. 

The numerical method described above, including boundary treatments, combining non-reflecting conditions, stretched grid buffer zone, and anchoring inflow variables to target values was developed and tested for jet aeroacoustics studies, see S. Ganesh\cite{SubramanianPHD}. And later successfully tested to understand jet instability modes on Hartmann whistle, see R.Varadharajan \textit{et al.}\cite{Varadharajan2016on}

\begin{figure}[htbp!]
\centering
\includegraphics[trim={0cm 2cm 0 2cm}, clip, width=1.0\textwidth]{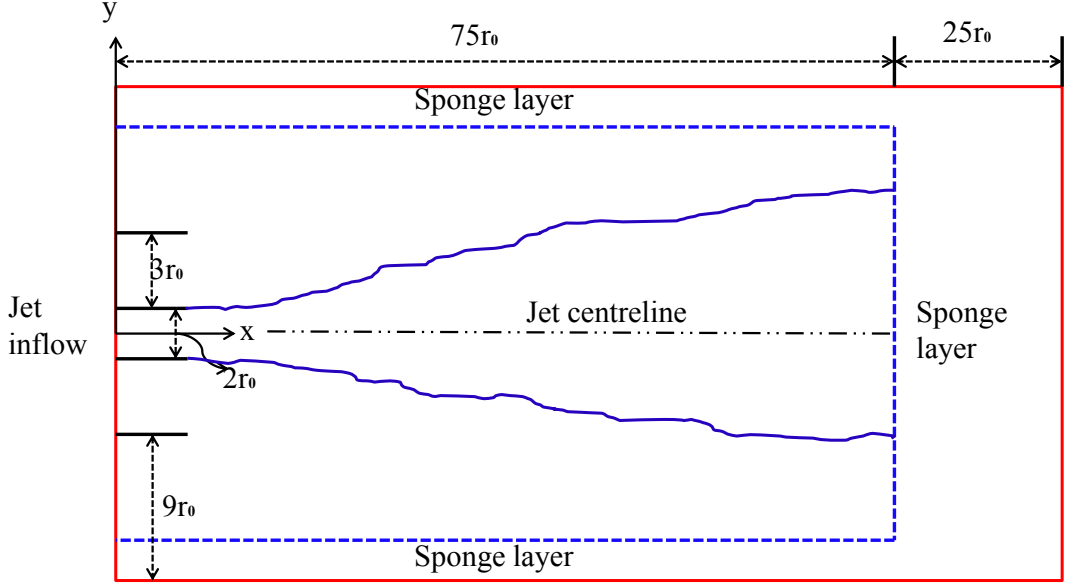}
\caption{Schematic of computational domain on the plane $z = 0$, in three-dimensional computational space. Sponge-layer is implemented to avoid spurious reflections from boundaries as shown in figure. }
\label{fig:comp_dom}
\end{figure}

\begin{figure*}[htbp!]
\centering
    \begin{subfigure}[t]{0.485\textwidth}
    \centering 
    \includegraphics[width=1.0\textwidth]{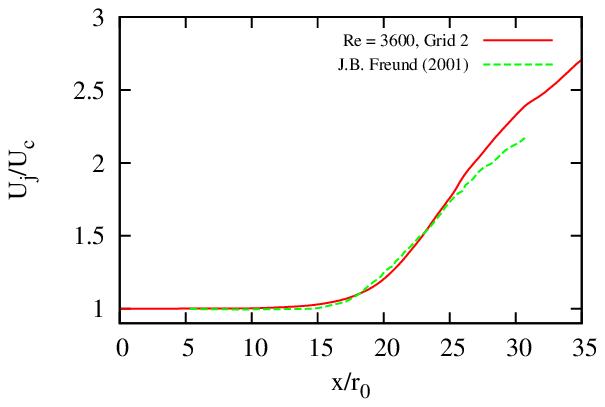}
    \caption{\small{M$=0.9$, Re$=3600$}}
    \label{fig:cdre3600}
    \end{subfigure}
    \begin{subfigure}[t]{0.485\textwidth}
    \centering 
    \includegraphics[width=1.0\textwidth]{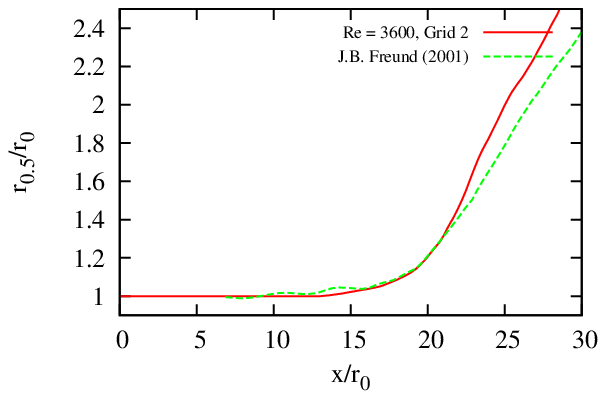}
    \caption{\small{M$=0.9$, Re$=3600$}}
    \label{fig:hwre3600}
    \end{subfigure} \\
    \begin{subfigure}[t]{0.485\textwidth}
    \centering 
    \includegraphics[width=1.0\textwidth]{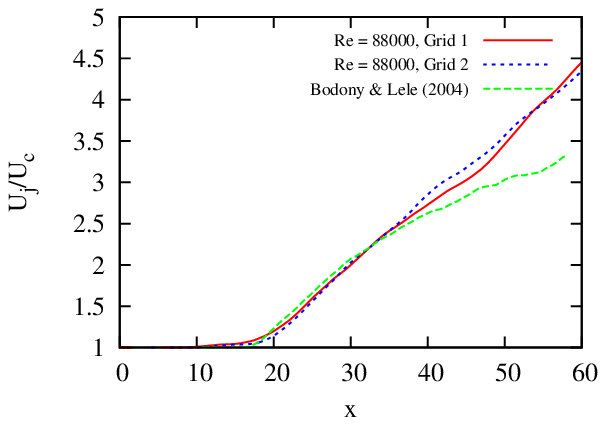}
    \caption{\small{M$=0.9$, Re$=88000$}}
    \label{fig:cdre88000}
    \end{subfigure} 
    \begin{subfigure}[t]{0.485\textwidth}
    \centering 
    \includegraphics[width=1.0\textwidth]{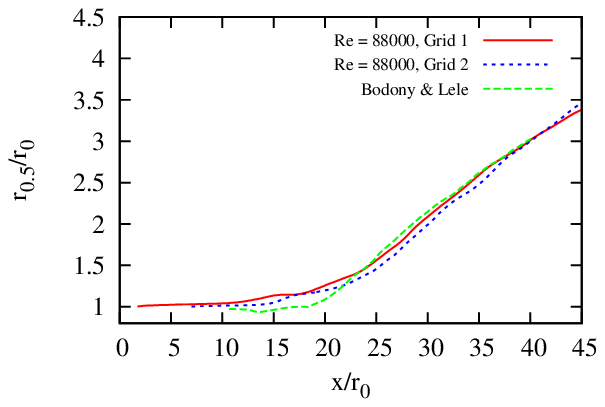}
    \caption{\small{M$=0.9$, Re$=88000$}}
    \label{fig:hwre88000}
    \end{subfigure} 
    \begin{subfigure}[t]{0.485\textwidth}
    \centering 
    \includegraphics[width=1.0\textwidth]{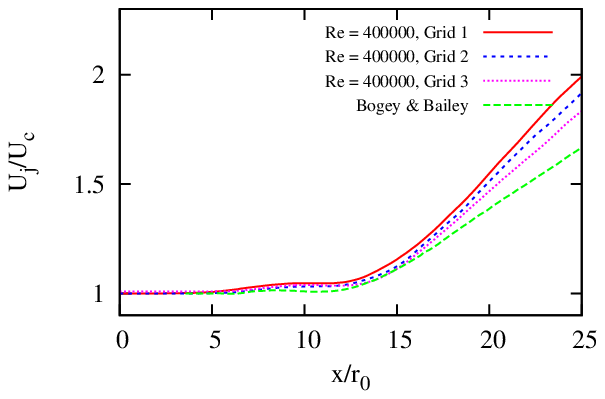}
    \caption{\small{M$=0.9$, Re$=400000$}}
    \label{fig:cdre400000}
    \end{subfigure}    
    \begin{subfigure}[t]{0.485\textwidth}
    \centering 
    \includegraphics[width=1.0\textwidth]{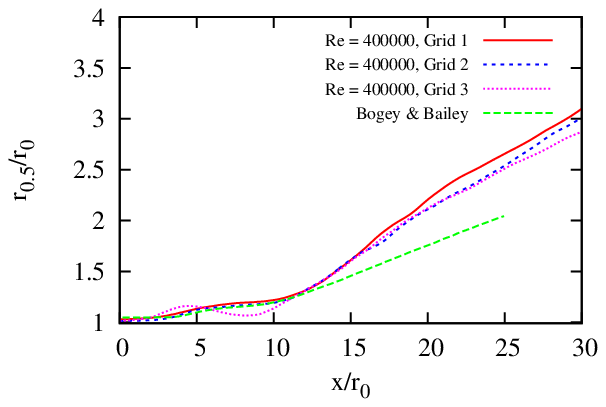}
    \caption{\small{M$=0.9$, Re$=400000$}}
    \label{fig:hwre400000}
    \end{subfigure}
\caption[Decay of centerline axial velocity \& streamwise variation of jet half-width radius.]{\small{Decay of centerline axial velocity \& streamwise variation of jet half-width radius. }}  
\label{fig:cdhwfull}
\end{figure*}

\begin{figure*}[htbp!]
\centering
    \begin{subfigure}{0.495\textwidth}
    \centering 
    \includegraphics[width=1.0\textwidth]{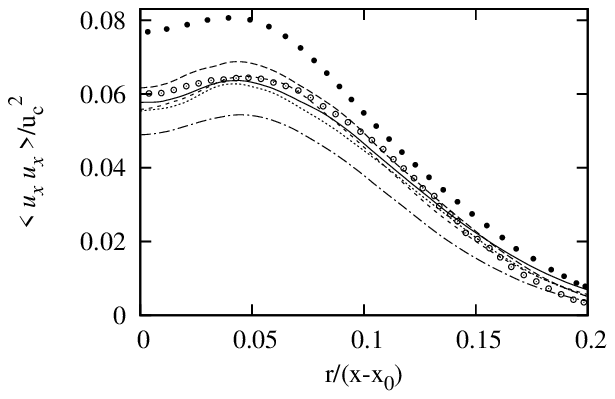}
    \label{fig:uxrms_ss}
    \end{subfigure}
    \begin{subfigure}{0.495\textwidth}
    \centering 
    \includegraphics[width=1.0\textwidth]{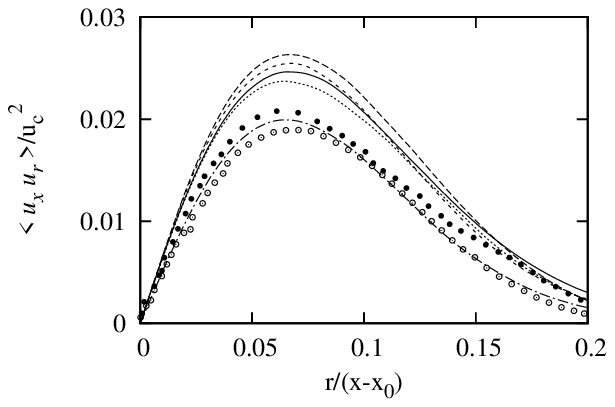}
    \label{fig:uxur_ss}
    \end{subfigure} \\
    \begin{subfigure}{0.495\textwidth}
    \centering 
    \includegraphics[width=1.0\textwidth]{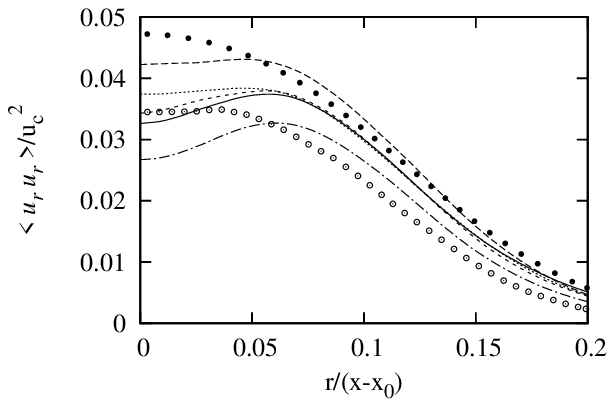}
    \label{fig:urrms_ss}
    \end{subfigure}
    \begin{subfigure}{0.495\textwidth}
    \centering 
    \includegraphics[width=1.0\textwidth]{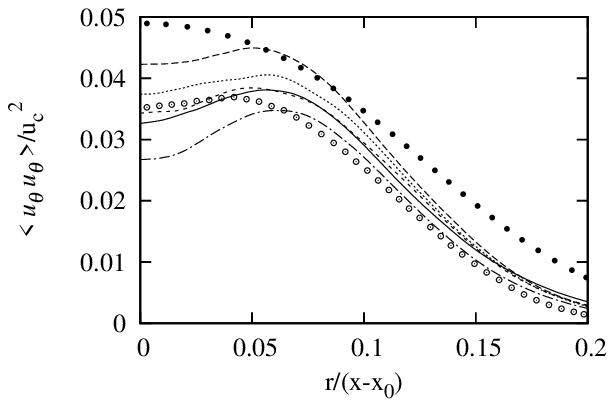}
    \label{fig:utrms_ss}
    \end{subfigure}   
\caption[Second moments of velocity]{\small{Second moments of velocity:  \tikz\draw[black,fill=black] (0,0) circle (.5ex); - Hussein \textit{et al.} 
(1994); \tikz\draw[black,fill=white] (0,0) circle (.5ex); - Panchapakesan \& Lumley (1993); (\ref{hwplot2}) - Re = 88000, Grid 1;
 (\ref{hwplot1}) - Re = 88000, Grid 2; (\ref{hwplot4}) - Re = 400000, Grid 1;
 (\ref{hwplot3}) - Re = 400000, Grid 2; (\ref{hwplot5}) - Re = 400000, Grid 3.}}  
\label{fig:ssprofiles}
\end{figure*}

\begin{figure*}[htbp!]
\centering
    \begin{subfigure}{0.49\textwidth}
    \centering 
    \includegraphics[trim={0 0 0 0.8cm},clip,width=1.0\textwidth]{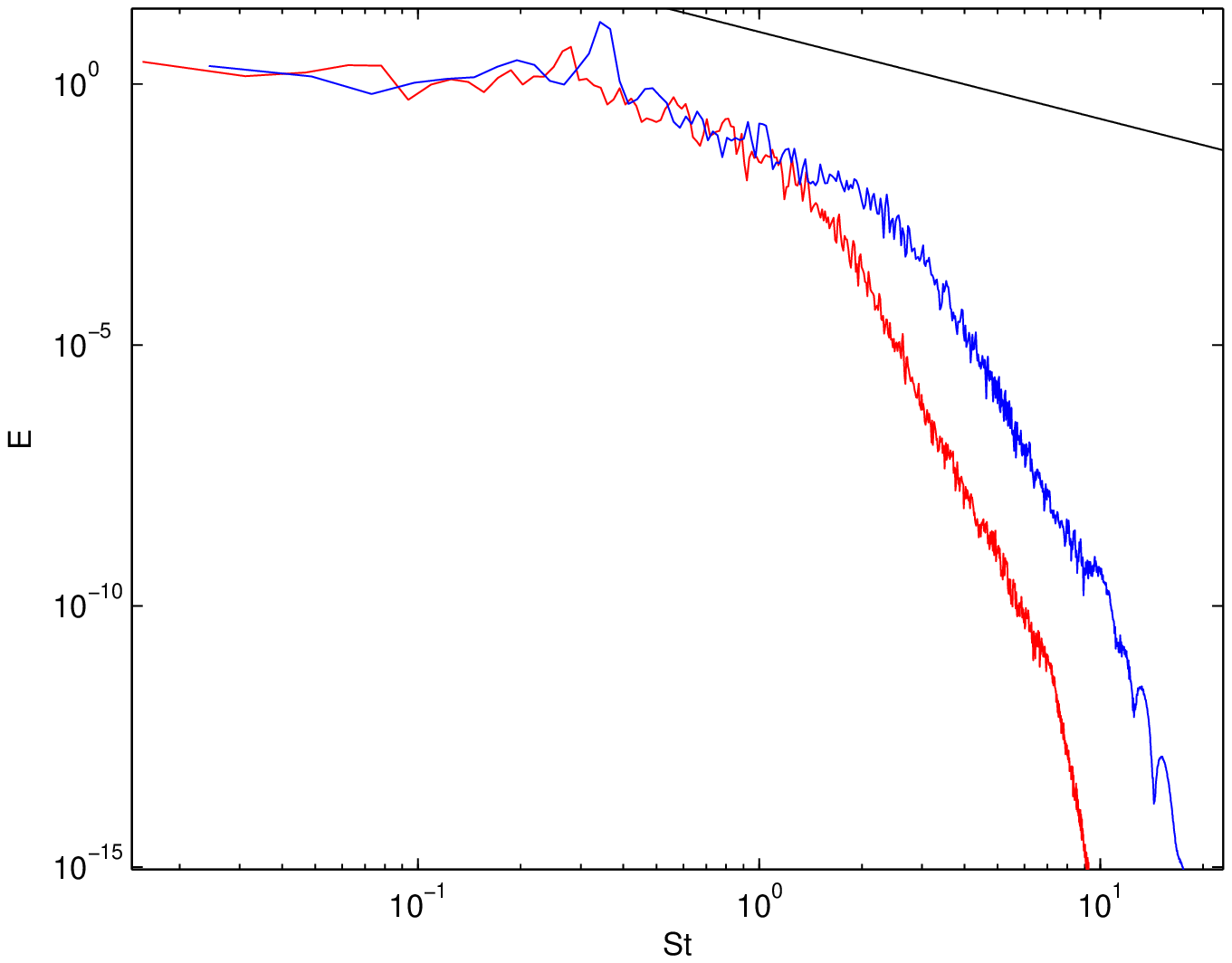}
    \caption[TKE spectrum of J2 jets]{\small{Re = 88000}}
    \label{fig:tkej2}
    \end{subfigure}
    \begin{subfigure}{0.49\textwidth}
    \centering 
    \includegraphics[trim={0 0 0 0.8cm},clip,width=1.0\textwidth]{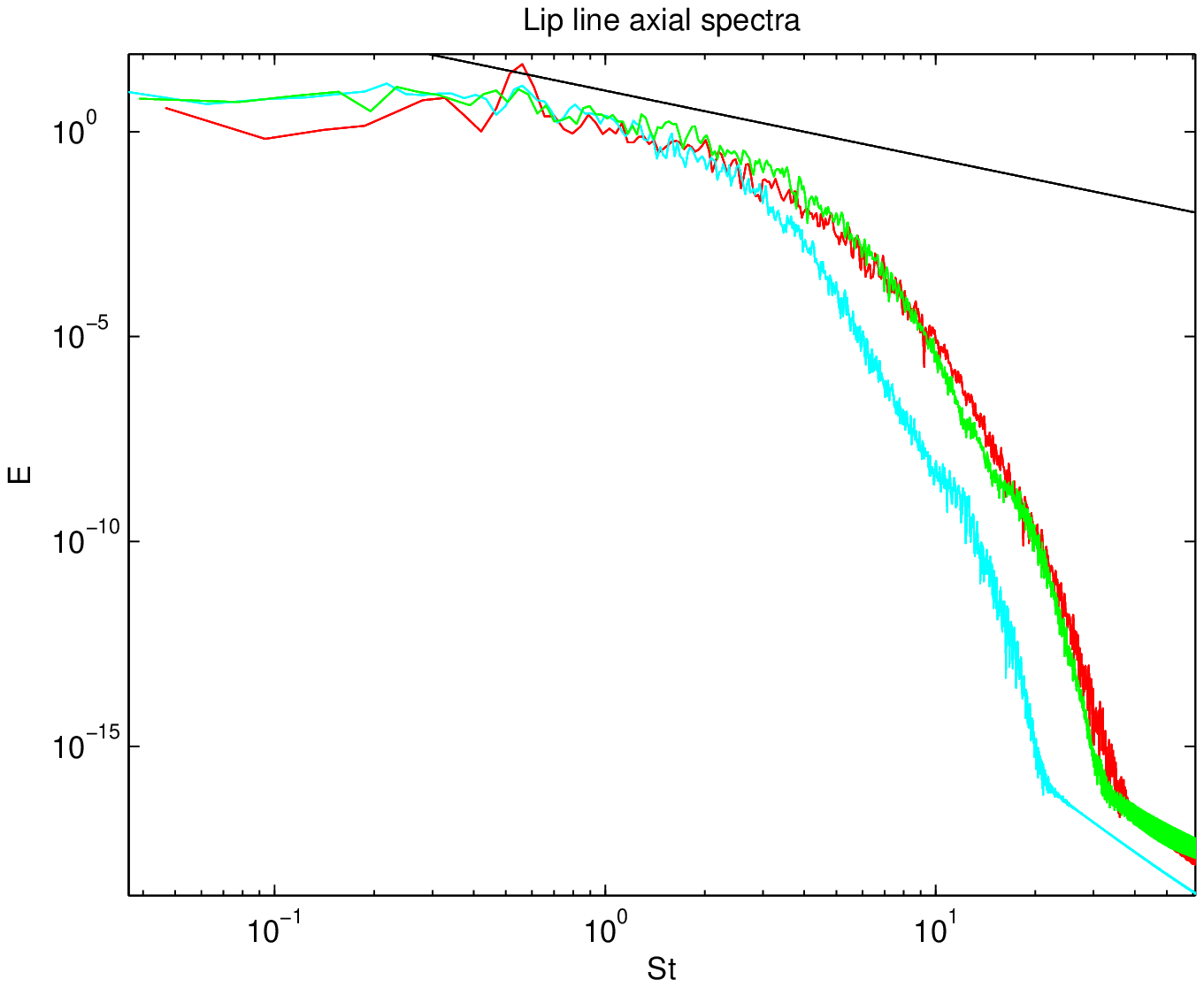}
    \caption[TKE spectrum of J2 jets]{\small{Re = 400000}}
    \label{fig:tkej3}
    \end{subfigure} 
\caption[Turbulent kinetic energy spectrum]{\small{Turbulent kinetic energy spectrum: Re = 88000: [(\ref{cline1})- Grid 1; (\ref{cline2}) - Grid 2.]
Re = 400000: [(\ref{cline3})- Grid 1; (\ref{cline4}) - Grid 2; (\ref{cline1}) - Grid 3.]; (\ref{hwplot1}) - $5/3$ slope line}}  
\label{fig:tkeprofiles}
\end{figure*}

\begin{figure*}[htbp!] 
\centering    
\includegraphics[width=0.7\textwidth]{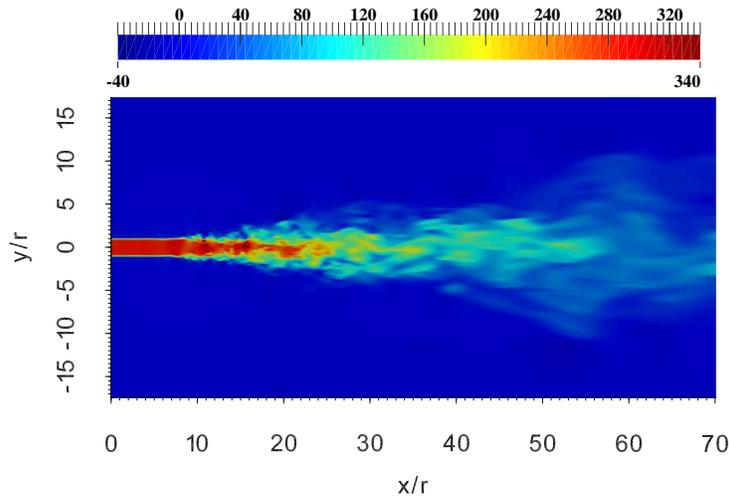}
\caption[Instantaneous contours of u-velocity]{Instantaneous contours of u-velocity at $xy$ plane, \ $z=0$. $Re = 4 \times 10^{5}$, showing potential-core collapse at $x = 14.5r_0$.}
\label{fig:uvel_re400000g3}
\end{figure*}

\begin{table*}
\caption{Details of round jet simulations}
\centering
\label{table:simulation_detail}
\begin{tabular}{l r c r r}
\toprule
\multirow{2}{*}{Simulation} & \multicolumn{2}{c}{jet Details} & \multicolumn{2}{c}{Parameters} \\ 
\cmidrule{2-5}
  & Reynolds Number & Grid (No.of.Div. in jet)  & $B_u$ & $x_{jc}/r_0$  \\ 
\midrule
J1G2a & 3600 & $386 \times 201 \times 201$ (10) & 5.96  & 42.4 \\

J2G1 & 88000 & $226 \times 171 \times 171$ (08) & 6.37  & 20.2 \\

J2G2 & 88000 & $386 \times 201 \times 201$ (10) & 6.28 & 19.8 \\

J3G1 & 400000 & $226 \times 171 \times 171$ (08) & 7.40 & 14.8 \\

J3G2 & 400000 & $386 \times 201 \times 201$ (10) & 6.60 & 14.5 \\

J3G3 & 400000 & $291 \times 241 \times 241$ (12) & 6.55 & 14.2 \\ 

Lumley, 1993 & 11000 & N.A.(Experiment) & 6.06 & -- \\ 

Freund, 2001 & 3600 & $640 \times 250 \times 160$(55) & 5.80 & 14.0 \\ 

\bottomrule
\end{tabular}
\end{table*}

\section{Jet inflow conditions}
\label{S:3}
Round jet with thin shear layer is provided as inflow. Velocity profile of jet is defined by top-hat profile
given below,

\begin{equation}
 U = \frac{U_j}{2} + \frac{U_j}{2} \ \tanh\bigg{(}\frac{r_0 - r}{2 \theta} \bigg{)}
\end{equation}

The above specified nozzle inflow condition represents jet with a thin laminar annular shear layer with radius $r_0$ and jet velocity $U_j$.
Momentum thickness, $\theta = r_0/20$, is used for simulations.
Mean inlet pressure is constant and maintained same as the ambient chamber pressure, $p_0 = 1823.85 \ Pa$. Inlet
density is determined using the Crocco-Busemann relationship. Ambient density, $\rho_0 = 0.022 \ kg/m^{3}$
and inflow density, $\rho_{in} = 0.02564 \ kg/m^{3}$ are used.
Mach Number of simulated jets are set to constant value of 0.9. 
Random fluctuations are added to the stream-wise components at inflow to induce turbulence. The fluctuations are calculated as follows,

\begin{equation}
 u' (\phi, t) = A \Bigg{[} \sum_{i=1}^{6} \sin (i\phi+\alpha_i+\psi_it) + \sum_{i=7}^{8} \sin(\psi_it)\Bigg{]}_{(x = 0)}
\end{equation}
\begin{equation}
A = \frac{U_j}{1000} \ e^{\big{(}-128(\frac{r - R}{R})^{2}\big{)}}
\end{equation}
where $\phi$ is the azimuthal angle. $\alpha_i$ and $\psi_i$ are random numbers to generate random phase and frequencies.
R = 0.875$r_0$ is used, providing with a gaussian function centered around R near nozzle lip. The perturbation
$u'$ is added at each step of RK2. The peak rms of the stream-wise velocity fluctuations are under 2\% for 
all cases.

\section[Computational domain]{Computational domain}
\label{S:4}

Computation domain used for simulations, are cartesian grids with \textit{x-axis} along streamwise direction, \textit{y,z axis} along cross-stream direction and origin
located at jet centerline, is schematically represented in Figure \ref{fig:comp_dom}. Three different grids with similar configuration are used as
computational domain for all numerical simulations. 
6 Simulations of Mach 0.9, Re = 3600, 88000, $4 \times 10^5$ round jets, in three different grids, are performed to study the effect of Reynolds number.

Numerical grid with 16 uniform divisions in jet diameter, both in y and z direction, is called as Grid 1 (G1) throughout the article. 
Grid 1 is used to simulate two cases of jet with Re $= 88000, 4 \times 10^5;$ \. 
From $ y = \pm 4r_0$ and $z = \pm 4r_0$, $2$\% stretching in local $\Delta$y is used.
Buffer region with $5$ \% stretching is used above $x = 75r_0$.

Numerical grid with 20 uniform divisions in jet diameter is called as Grid 2 (G2) throughout the article. 
Grid 2 is used to simulate three cases of jet with Re $= 3600$ (Grid 2a is used), $88000, 4 \times 10^5;$ \. 
From $ y = \pm4r_0$ and $z = \pm4r_0$, $2$ \% stretching in local $\Delta$y is used. Buffer region with $5$ \% stretching is used above $x = 75r_0$.
Grid 2a is a variant of Grid 2, where the compression in the stream-wise direction is started at $x = 40r_0$ (which was the expected breakdown length for $Re = 3600$ case).

Numerical grid with 24 uniform divisions in jet diameter is called as Grid 3 (G3) throughout the article. 
Grid 3 is used to simulate single case of jet with Re $= 4 \times 10^5;$ \. From $ y = \pm 4r_0$ and $z = \pm 4r_0$, $2$ \% stretching in local $\Delta$y is used. 
Buffer region with $5$ \% stretching is used above $x = 75r_0$.

All three grids are compressed from $x =6r_0$ and stretched after $x = 12r_0$ approximately. 
Grid is compressed to capture finer length scales in the near break down region. 
The region of compression is chosen based on expected break down point from previous numerical experiments. 
Compression and Stretching parameters $r_c = 0.985$ and $r_s = 1.015$ represent the common ratio of compression and stretching respectively for all three grids.
Grids described above are all developed as a function of jet radius $r_0$.

\section[Simulation results]{Simulation results}
\label{S:5}

\subsection[Mean flow]{Mean flow}

In the numerical computations, there is an ambiguity in the definition of effective nozzle location as the computational domain doesn't include the nozzle. 
Therefore, virtual origin values of these simulations are matched to compare the simulation results with experimental and other computational works.
Virtual origin values are obtained from the jet half-width radius. A linear fit for half-width radius is found in the self-similar region.
This linear fit is extrapolated to find the virtual origin of the jets.

Self-Similarity \cite{pope2001turbulent}\cite{benzi1993extended} states, downstream of jet breakdown and potential core collapse, the centerline velocity decays according to, 

\begin{align}
 \frac{U_c}{U_j} = 2B_u \frac{r_0}{x-x_0} 
\end{align}

\noindent where $B_u$ is a measure of center-line axial-velocity decay rate. For the cases of simulations performed center-line axial velocity decay and
Half-width radius are presented in Figure [\ref{fig:cdhwfull}]. 
From Table \ref{table:simulation_detail}, it can be observed that jets with higher Reynolds number has higer values of $B_u$. This shows that increase in Reynolds number causes decreased axial-velocity decay rate for turbulent round jets. It should also be noted that there isn't any significant variation of spread rate for different Reynolds numbers(Can be observed from jet halfwidth plots($r_{0.5}/r_0$ $vs.$ $x/r_0$) in Figure \ref{fig:cdhwfull}). In Figure \ref{fig:cdhwfull}(e,f) the slight deviation can be attributed to use of Cylindrical geometry by Bogey \& Bailey\cite{Bogey2006}. 

\subsection[Turbulence]{Turbulence}
Downstream of the jet breakdown, jets become self similar. In the self similar region, the 
second moments of velocity are used to characterize the turbulence. Figure [\ref{fig:ssprofiles}] shows that post-transition turbulence intensities decreases as Reynolds number increases.
It can be observed that peak turbulence intensities are 33\% lower for Re = 400000 when compared to Re = 88000 jets.
This is similar to the observations of Xu \textit{et al.}\cite{xu2013effects} for turbulent plane jets.

For all the simulations performed the Reynolds stress behavior is found to lie
mostly within the profiles measured by Panchapakesan \& Lumley \cite{panchapakesan1993turbulence} and
Hussein \textit{et al.} \cite{hussein1994velocity}. Slight deviation observed in near centerline region for some of the 
simulation is because the profiles were averaged in same range $ x \ge 20r_0 $, for all simulations, where some of the
jets simulated are in transition or near-transition zone. Turbulence intensity profiles of J1G2a are not presented as the turbulence regime simulated for the case is small, and averaging in that span will not produce accurate results.\\

\subsection[Energy spectra]{Energy spectra}
Turbulent kinetic energy spectra, for simulations performed, are presented in Figure: [\ref{fig:tkeprofiles}]. Energy spectra is broad banded with 
the smallest scales having approximately $10^{6}$ times less energy than greatest scale for all simulations. Turbulent kinetic
energy of the simulated jets peaks around Strouhal Number, St $ = 0.3 - 0.4$. Figure: [\ref{fig:tkej2} \& \ref{fig:tkej3}]
shows that on grid refinement finer scales are captured accurately. Figure: [\ref{fig:tkeprofiles}] shows energy content of 
a fixed length-scale (Strouhal Number) increases with increase in Reynolds number. 
This is observed for smaller length-scales or higher strouhal numbers (St $\ge$ 0.5), whereas the 
energy content in the highest length-scales or lower strouhal numbers remain independent of Reynolds number. \\

\subsection[Flow organisation]{Flow organisation}
Axial velocity contour of simulated jet is presented in Figure: [\ref{fig:uvel_re400000g3}]. Except for the 
differences in transition length, flow organisation remains the same for all simulations. Potential core 
collapse is observed at a range of $x = 14r_0$ - $20r_0$ on an average for the simulations performed, with exception to Re $=3600$ case. (where
the potential core collapse is observed at $x = 39r_0$). 
It is observed that high Reynolds number jets have early transition (potential core collapse) for the same level of inflow turbulence.

\section[Summary]{Summary}
An early break-down to turbulence, decreased axial-velocity decay rates and decreased levels of turbulence intensities are observed with increase in Reynolds numbers for turbulent round jets.
Eventhough more accurate numerical simulations for a range of Reynolds number are required to understand the effects, it could be observed that increased smaller length scales play a significant role in determining decay rates. 
Jet spread-rate though observed to remain almost independent of Reynolds number, it could be an artifact of cartesian grids inability to represent the physics. 
Simulations on a cylindrical grid, might reveal dependence of spread-rate on Reynolds number.

\section*{Acknowledgements}
Author acknowledges the fruitful discussions with Dr. Joseph Mathew, Indian Institute of Science, Bangalore. Author is thankful to Dr. Subramanian Ganesh, General Electric, India for providing the basic computational algorithm. Computations were carried during author's stay at Indian Institute of Science, India, as a part of his graduate research on jet instability modes of Hartmann whistle. Analysis and findings published in this article were made after author moved to his new position at Wageningen University, The Netherlands.


%

%
\pagebreak
%
%
%
\appendix\label{section:references}
%
%
\bibliography{ascexmpl}
%
%
%

\begin{tikzpicture}
    \begin{axis}[hide axis]
        \addplot [color=red,solid,line width= 1.0 pt,forget plot]
        (0,0);\label{cline1}
        \addplot [color=blue,solid,line width= 1.0 pt,forget plot]
        (0,0);\label{cline2}
        \addplot [color=cyan,solid,line width= 1.0 pt,forget plot]
        (0,0);\label{cline3}
        \addplot [color=green, solid, line width= 1.0 pt, forget plot]
        (0,0);\label{cline4}
    \end{axis}
\end{tikzpicture}%

\begin{tikzpicture}
    \begin{axis}[hide axis]
        \addplot [
        color=black,
        solid,
        line width= 1.0 pt,
        forget plot
        ]
        (0,0);\label{hwplot1}
        \addplot [
        color=black,
        densely dashed,
        line width=1.0pt,
        forget plot
        ]
        (0,0);\label{hwplot2}
        \addplot [
        color=black,
        dotted,
        line width=1.0pt,
        forget plot
        ]
        (0,0);\label{hwplot3}
        \addplot [
        color=black,
        loosely dashed,
        line width=1.0pt,
        forget plot
        ]
        (0,0);\label{hwplot4}
        \addplot [
        color=black,
        dashdotted,
        line width=1.0pt,
        forget plot
        ]
        (0,0);\label{hwplot5}
    \end{axis}
\end{tikzpicture}%

\end{document}